\documentclass[twocolumn,showkeys,preprintnumbers,amsmath,amssymb]{revtex4}

\usepackage{graphicx}
\usepackage{dcolumn}
\usepackage{bm}

\begin{document}
\preprint{\it Astronomy Letters, 2014, Vol. 40, Nos. 2 –- 3, pp. 75
–- 85\\}

\title{Emission from the Galaxy NGC 1275 at High \\
and Very High Energies and Its Origin}

\author{V. G. Sinitsyna and V. Yu. Sinitsyna\vspace{4mm}}
\email{sinits@sci.lebedev.ru}

 \affiliation{\\ Lebedev Physical Institute, Russian Academy of Sciences, \\ Leninskii pr. 53, Moscow, 119991 Russia \\}

\date{Received April 12, 2013; in final form, October 13, 2013}

\begin{abstract}
\vspace{0.5cm} {\bf Abstract --} The Seyfert galaxy NGC 1275 is the
central, dominant galaxy in the Perseus cluster of galaxies. NGC1275
is known as a powerful source of radio and X-ray emission. The
well-known extragalactic object NGC 1275 has been observed by the
SHALON high-altitude mirror Cherenkov telescopes within the
framework of long-term studies of metagalactic gamma-ray sources. In
1996, the SHALON observations revealed a new metagalactic source of
very high energy gamma-ray emission coincident in its coordinates
with the galaxy NGC 1275. Having analyzed the SHALON data, we have
determined such characteristics of NGC 1275 as the spectral energy
distributions and images at energies $>$800 GeV for the first time.
The results obtained at high and very high energies are needed for
understanding the emission generation processes in an entire wide
energy range.\vspace{0.5cm}
\end{abstract}


\maketitle

\section*{INTRODUCTION}

The cluster of galaxies in Perseus is one of the best-studied
clusters owing to its relative proximity (its distance ¡«100 Mpc or
redshift z = 0.0179) and brightness. Clusters of galaxies have long
been considered as possible candidates for the sources of TeV
gamma-rays emitted by protons and electrons accelerated at
large-scale shocks or by a galactic wind or active galactic nuclei
(Dennison 1980; Houston et al. 1984; Colafrancesco and Blasi 1998;
Sarazin 1999; Miniati et al. 2001; Timokhin et al. 2004;
Colafrancesco and Marchegiani 2009). The dominant galaxy in the
Perseus cluster is NGC 1275.

\section*{NGC 1275}

The galaxy NGC 1275 has been classified in various ways, for
example, as a Seyfert 1.5 galaxy, because broad emission lines were
detected in its spectrum at optical wavelengths (Veron-Cetty and
Veron 1998). However, within the unified model of active galactic
nuclei (AGNs) (Fanaroff and Riley 1974; Urry and Padovani 1995), it
also belongs to the main class of BL Lacertae objects owing to its
large and fast flux variability and polarization (Angel and Stockman
1980). It should be noted that there is evidence that the mentioned
AGN unification scheme can be even simplified (Kharb et al. 2010).

NGC 1275 is a powerful source of radio and X-ray emission. In the
radio band, the object found in NGC 1275, also known as Perseus A
and 3C 84, has a powerful and compact core that has been well
studied with VLBI (Vermeulen et al. 1994; Taylor and Vermeulen 1996;
Asada et al. 2006). NGC 1275 is extremely bright in the radio band
and was classified as an FR I radio galaxy; it has a prominent
structure that consists of a compact central source and an extended
jet (Vermeulen et al. 1994; Asada et al. 2009). Having a
supermassive black hole (with a mass of $3.4\times10^8 M_{\odot} $)
at its center (Wilman et al. 2005), NGC 1275 also exhibits jet
precession, which can be interpreted as a possible manifestation of
the fact that NGC 1275 is the result of a merger between two
galaxies (Liu and Chen 2007). The radio emission extends to great
distances and shows a clear interaction with the gas inside the
Perseus cluster of galaxies. ROSAT (B\"ohringer et al. 1993) and,
subsequently, Chandra (Fabian et al. 2006) observations revealed
cavities in the gas located inside the cluster, whose presence
suggests that the jets from 3C 84 sweep up numerous "bubbles" in the
atmosphere of the Perseus cluster (Fig. 1).
\begin{figure}
\begin{center}
\includegraphics{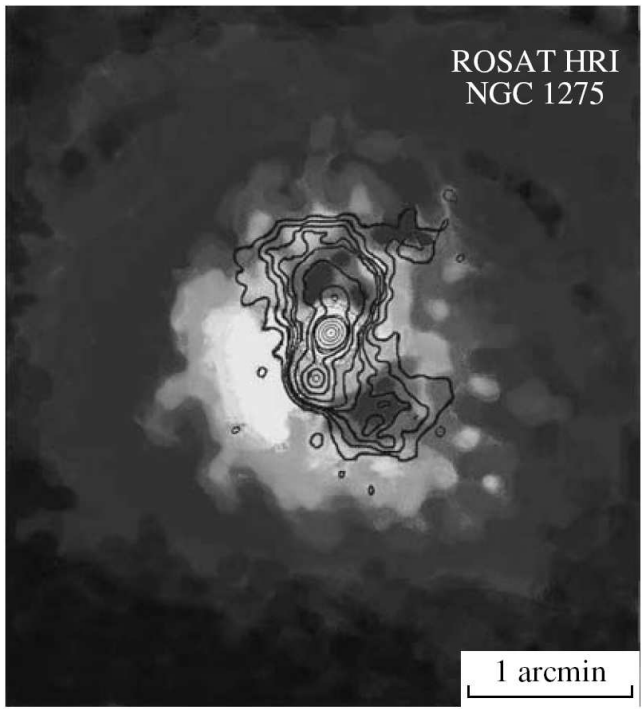}
\end{center}
\caption{ROSAT X-ray (0.1 -– 2.4 keV) image of NGC 1275 (B\"ohringer
et al. 1993). The contours represent the source's radio structure
from VLA radio observations. The radio and X-ray emission maxima
coincide with the active galactic nucleus NGC 1275, while the X-ray
emission disappears almost completely near the bright areas of the
radio components.}
\end{figure}

The galaxy NGC 1275 surrounded by extended filamentary structures
historically aroused great interest owing to both its position at
the center of the Perseus cluster and its possible "feedback" role
(Gallagher 2009). Evidence for the "feedback" role of NGC 1275 can
be obtained from ROSAT and Chandra observations, which reveal shells
of hot gas and cavities that spatially coincide with the radio
structures (Fig. 1) extending from the central, active part of the
AGN. NGC 1275 also arouses interest owing to its close proximity to
the Earth at redshift z = 0.0179 (Strauss et al. 1992), making it
possible to study the physics of relativistic jets.
\begin{figure*}
\begin{center}
\includegraphics[width=4.2in]{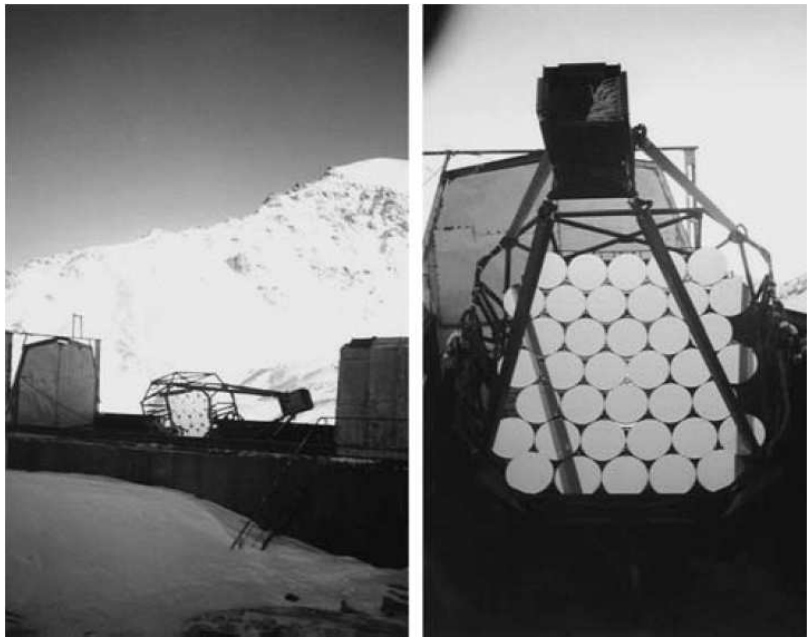}
\end{center}
\caption{SHALON-1 mirror Cherenkov telescope in the observatory.}
\end{figure*}

Upper limits on the gamma-ray emission from the Perseus cluster of
galaxies and its central galaxy NGC 1275 were obtained in various
experiments onboard satellites. The first observations were
performed with the COS-B telescope from 1975 to 1979 (Strong et al.
1982) and then with the EGRET facility in 1995 (Thompson et al.
1995).

At very high energies, upper limits were obtained in different years
in ground-based experiments, such as the large-area scintillation
Tibet Array at E $>$ 3 TeV (1999) (Amenomori 1999), and at the
Cherenkov telescopes Whipple (2006) (Perkins et al. 2006) at
energies $>$400 GeV, MAGIC (2009) (Al\'eksic et al. 2010) at E $>$
100 GeV, and Veritas (2009) (Acciari et al. 2009) at E $>$ 188 GeV.
Recently, NGC 1275 was recorded at high energies, 100 MeV – 300 GeV,
by the Fermi LAT satellite telescope (Abdo et al. 2009). To
understand the emission generation processes in the entire energy
range, the spectral energy distribution should be extended up to
very high energies.

In this paper, we describe both the results of fifteen-year-long
studies for the AGN NGC 1275 obtained at very high energies with the
SHALON mirror Cherenkov telescope and the experimental approach to
searching for the gamma-ray emission mechanisms in clusters of
galaxies and AGNs located in the cluster atmosphere.

\section*{THE SHALON MIRROR CHERENKOV \\ TELESCOPES}

The SHALON mirror gamma-ray telescopes of the Lebedev Physical
Institute, the Russian Academy of Sciences, with which the data
presented here were obtained are the only operating gamma-ray
telescopes in the Russian Federation and one of the five telescopic
facilities in the world that are currently performing systematic
observations of local gamma-ray sources at TeV energies. Methodical
experiments and observations at the first SHALON mirror Cherenkov
telescope were begun twenty years ago. The SHALON experiment and its
main characteristics are described in Sinitsyna (1995, 1996, 1997,
2000, 2006), Sinitsyna et al. (1998, 2003, 2007, 2009), Nikolsky and
Sinitsyna (1989, 2004), and V.G. Sinitsyna and V.Yu. Sinitsyna
(2011a, 2011b). The telescope's parameters, the observing technique,
and the selection criteria of gamma-ray showers from background
cosmic-ray showers are also described in Nikolsky and Sinitsyna
(1989), Sinitsyna (1995, 1996), and V.G. Sinitsyna and V.Yu.
Sinitsyna (2011a, 2011b).

The SHALON ALATOO mirror Cherenkov telescope system (Fig. 2) is
designed to observe gamma rays from local sources in the energy
range from 800 GeV to 100 TeV. The SHALON gamma-ray telescopes are
located at an altitude of 3340 m above sea level each of which has a
composite mirror with an area of 11.2 m$^2$. The detector array
consisting of 144 FEU-85 photomultiplier tubes assembled into a
square array and mounted at the mirror focus has characteristics
sufficient to record information about the shower structure in the
energy range under consideration. The detector has the largest field
of view in the world, $> 8^\circ$. This allows one to monitor the
background from charged cosmic-ray particles and the atmospheric
transparency continuously during observations and expands the area
of observation and, hence, the efficiency of observations (Nikolsky
and Sinitsyna 1989; Sinitsyna 1995, 1996). The technique for
simultaneously obtaining information about the cosmic-ray background
and the showers initiated by gamma rays is unique and has been
applied in the SHALON experiment from the very beginning of its
operation (Nikolsky and Sinitsyna 1989; Sinitsyna 1995, 1996). This
technique serves to increase the useful source tracking time and,
what is particularly important, such source and background
observation conditions as the thickness and state of the atmosphere
remain the same. This method is inaccessible to other gamma-ray
astronomical experiments, because the telescopes used in the world
have a smaller field of view. In addition, the wide field of view
allows recording the off-center showers arriving at distances of
more than 30m from the telescope axis completely and almost without
any distortions; they account for more 90\% of all the showers
recorded by the telescope. During a primary analysis, the primary
particle arrival direction is determined with an accuracy up to
$\lesssim 0.1^\circ$. The subsequent analysis specially developed
for the SHALON telescopes and based on Tikhonov's regularization
method (Tikhonov and Arsenin 1979; Goncharsky et al. 1985) improves
the accuracy to a value of less than $0.01^\circ$.

\begin{figure*}
\begin{center}
\includegraphics[width=2.9in]{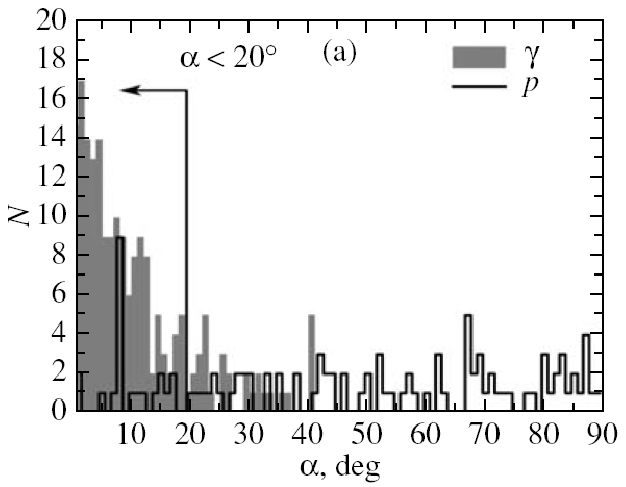}
\includegraphics[width=2.6in]{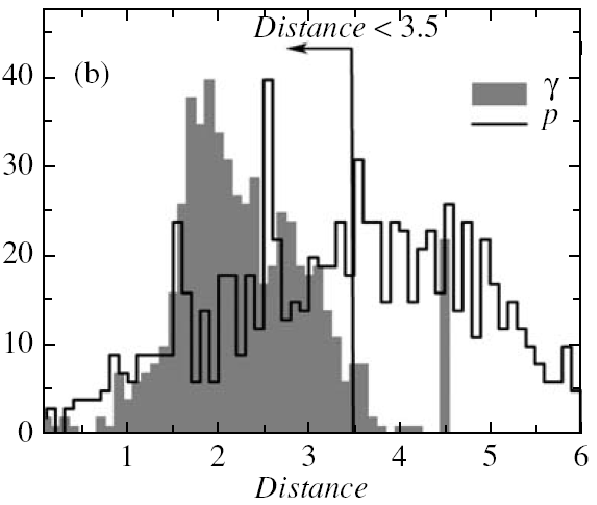}
\end{center}
\caption{Distributions of the parameters $\alpha$ (a) and $Distance$
(b) for gamma-ray showers and background showers (black contour).}
\end{figure*}

The main difficulty in detecting and investigating very high energy
gamma-ray sources is the presence of a significant (a factor of 1000
larger) background of cosmic rays producing the Cherenkov flashes in
the Earth's atmosphere that are hard to distinguish from the flashes
produced by gamma rays. Therefore, the determination of the criteria
for selecting gamma-ray showers from background cosmic-ray showers
(Sinitsyna 1996; V.G. Sinitsyna and V.Yu. Sinitsyna 2011b) is an
important stage of gamma-ray astronomical experiments.

The distributions of the shower image parameters used in the SHALON
experiment as a selection criteria optimized from the Monte Carlo
simulations are presented in Figs. 3 –- 5 below (Sinitsyna 1996).
The simulation correctness is confirmed by the distributions of
shower image parameters both for background of cosmic rays and for
gamma-rays from the point sources obtained experimentally in
observations (Sinitsyna 2003; V.G. Sinitsyna and V.Yu. Sinitsyna
2011b).
\begin{figure}[b!]
\begin{center}
\includegraphics[width=3.2in]{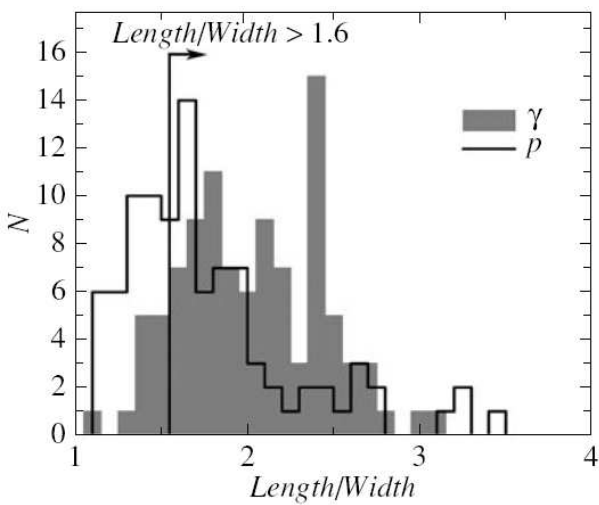}
\end{center}
\caption{Distribution of the parameter $Length/Width$ for gamma-ray
showers and background showers (black contour).}
\end{figure}

The Cherenkov radiation from an extensive air shower (EAS) falls on
the mirrors and, being reflected from them, illuminates some of the
photomultiplier tubes in the detector array (V.G. Sinitsyna and
V.Yu. Sinitsyna 2011b). The image of the Cherenkov light from a
shower in the detector plane is generally an elliptical light spot
with a central peak. Analyzing the angular and lateral distributions
of the shower Cherenkov light is of great interest for the selection
of showers of primary gamma-rays. The so-called Hillas parameters
(Hillas 1985) : $\alpha$, $Length$, $Width$, and $Distance$ are used
for these purposes in gamma-ray astronomy and optimized for each
experiment individually. The distributions of $\alpha$ (degrees) and
$Distance$ (pixels) for gamma-ray showers and background showers in
the SHALON experiment are presented in Fig. 3. About 72\% of the
background events can be rejected by applying the constraint $\alpha
< $ 20. Taking $Distance < $ 3.5 pixels as a criterion, we remove
$\sim$ 50\% of the background cosmic-ray showers.

\begin{figure*}
\begin{center}
\includegraphics[width=2.9in]{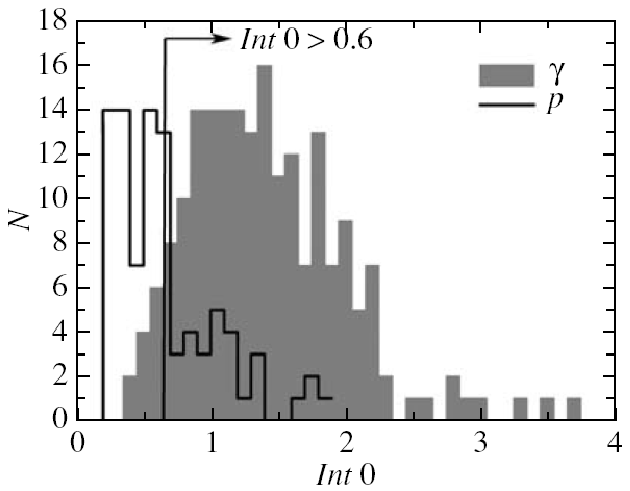}
\includegraphics[width=2.8in]{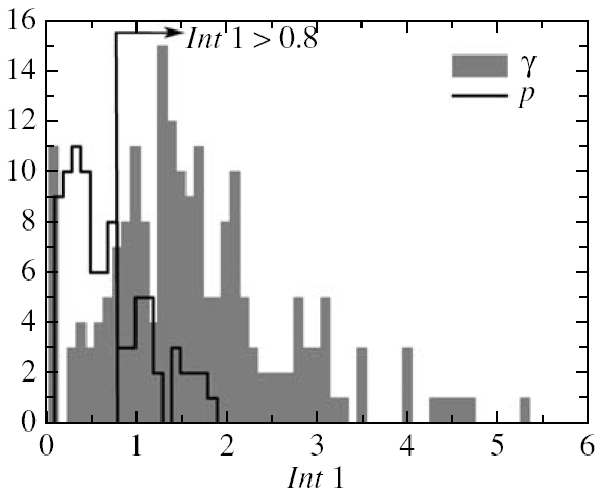}
\end{center}
\caption{Distributions of the parameters $Int0$ (left) and $Int1$
(right) for gamma-ray showers and background showers (black
contour).}
\end{figure*}
\begin{figure*}
\begin{center}
\includegraphics[width=2.9in]{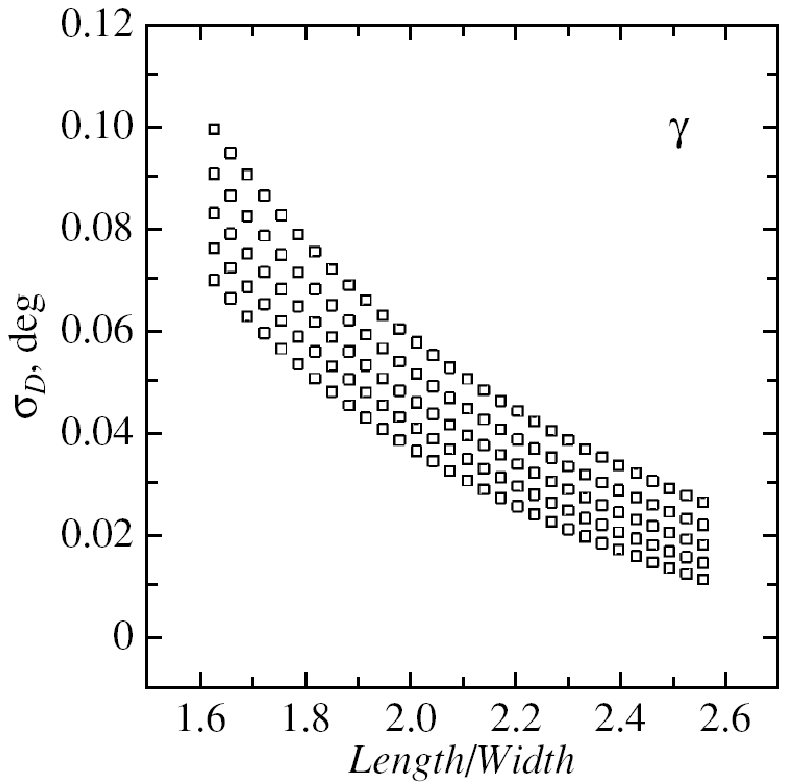}
\includegraphics[width=2.87in]{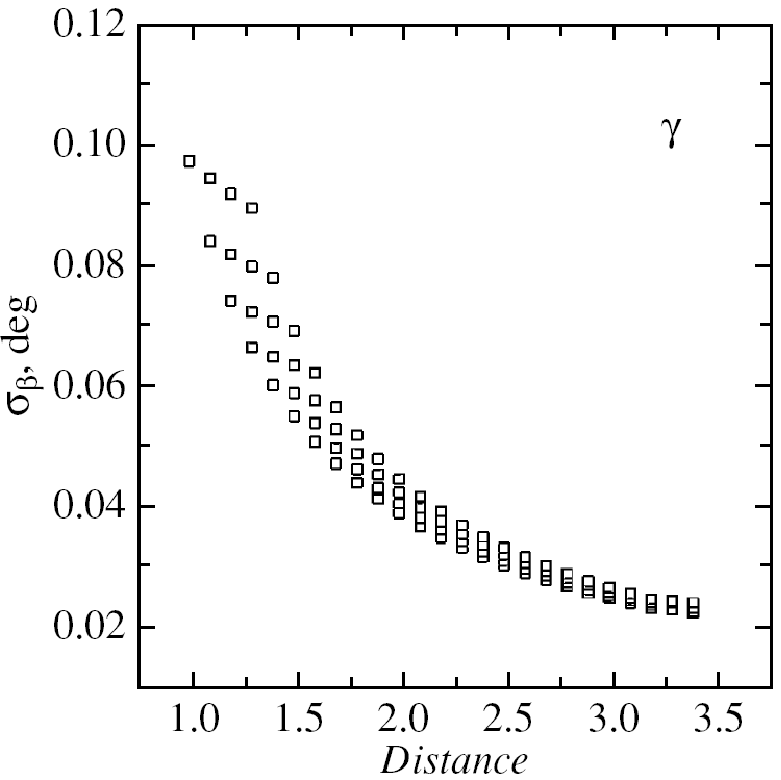}
\end{center}
\caption{Accuracies of the determination of the coordinates $D$
(left) and $\beta$ (right) for the point of the shower source on the
light receiver plane in the shower energy range 1 –- 100 TeV.}
\end{figure*}

Analysis of the distribution of $Length$ and $Width$ shower image
parameters showed that the mean values of these parameters for
gamma-ray and proton showers are, in principle, identical and,
hence, they cannot serve as a selection criterion. However, the mean
ratio $Length/Width$ is different for different types of showers and
is stable against a change in energy. Taking $Length/Width >$ 1.6 as
a criterion, we thus reject $\sim$ 49\% of the background showers,
leaving the number of gamma-ray showers intact (see Fig. 4).

Moreover, in addition to the standard criteria, it turned out to be
appropriate to construct one more characteristic of the light spot
which is based on the difference between the characteristics of
cascades of primary particles of a different nature. As a result,
selection criteria of the gamma-ray showers from the background ones
($Int0$, $Int1$) that are based on the difference between their
spatial characteristics were specially developed for the SHALON
gamma-ray telescope. For these purpose we use the difference between
their spatial characteristics, i.e., on the difference between the
characteristics of EASs of a different nature (Sinitsyna 1995, 1996;
V.G. Sinitsyna and V.Yu. Sinitsyna 2011b), if comparing the photon
fluxes within a small angle of less than $1^\circ$ around the
cascade axis with the flux within large angles of more than
$2^\circ$ (Nikolsky and Sinitsyna 1987) were specially developed for
the SHALON gamma-ray telescope. As a result, the parameters with
which the gamma-ray showers could be selected from the entire set of
available experimental light images were obtained. The pixel with
the maximum intensity (photoelectron density) $Intmax$ is chosen in
the array with a light image. Then, eight pixels surrounding the
maximum one are taken and the photoelectron intensities are summed
in them. The sum is designated as $Inteig$. The sum of the
photoelectron intensities in the remaining pixels (except for the
nine above-mentioned ones) is designated as $Intsur$ and the
following ratios are constructed: $Int0 = Intmax/Inteig$ and $Int1 =
Intmax/Intsur$. The constraints on these parameters for an efficient
selection of gamma-ray showers from cosmic-ray showers were derived
from the distributions of $Int0$ and $Int1$ (Fig. 5): $Int0 >$ 0.6
and $Int1 >$ 0.8. When the parameters $Int0$ and $Int1$ are applied,
$ $ 92 and 88\% of the background showers, respectively, are
rejected.

Analysis of the distributions showed that the following selection
criteria applied in the SHALON experiment ($\alpha < 20^\circ$,
$Length/Width > $ 1.6, $Int0 > $ 0.6, $Int1 >$ 0.8, $Distance <$ 3.5
pixels) reject 99.92\% of the background, while, according to our
estimations, the loss of gamma rays does not exceed 35\%, which is
taken into account in the subsequent analysis, but the contribution
of the background showers to the selected gamma-ray ones being no
more than 10\%.

The primary particle arrival direction and further construction of
the image are performed within a two-step procedure.

At the first step, the coordinates of the shower source position on
the light receiver plane for each of the showers selected according
to the described criteria are found. The shower image in the array
is characterized by an ellipse with the major axis that is the
projection of the shower axis onto the lightreceiver plane. The
gamma-ray shower source is on the extension of the major axis of
such an ellipse from the side of the shower maximum corresponding to
the cascade beginning. The distance from the centroid of the shower
image to the source's position, $D$, depends on the distance at
which the shower arrived and, as a result, on the shower elongation
by the parameter $Length/Width$. This dependence can be written as
$D = B\times[1-(Length/Width)^{-1}]$. The optimal proportionality
coefficient $B =$ 5.1 was chosen in such a way that the distribution
of shower arrival directions in angles was minimal in width and
centered at the source's position on the array.

The second coordinate $\beta$ is the inclination of the major axis
of the ellipse and can be defined via the parameter Distance. The
accuracies of the determination of each of these coordinates are
presented in Fig. 6 and are $0.07^\circ \pm 0.01^\circ$ and
$0.045^\circ \pm 0.01^\circ$, respectively, for the mean values of
$Length/Width$ and $Distance$ (Figs. 3 and 4).

At the second step, an additional analysis of the gamma-ray shower
source coordinates obtained at the first step is performed and the
gamma-ray intensity distribution in the source is found. Finding the
gamma-ray intensity distribution in the source $I(r)$ is reduced to
solving a Fredholm integral equation of the first kind: $ F(z) =
\int K(r, s)I(s)ds$ where $K(r, s)$ is the kernel function defined
as a Gaussian point spread function with the full widths half
magnitude $\sigma_D$ and $\sigma_\beta$ determined at the first
step. Then, we refine the observed distribution from Fig. 7a by
solving an integral equation (Goncharskii et al. 1985) under the
assumption that the solution is a smooth nonnegative upper-bounded
function, as the source's angular sizes were limited, and the
solution was definitely within the region determined at the first
step. Figure 7b presents the corrected gamma-ray intensity
distribution $I(r)$, where $r(SE)$ is the distance to the southeast
from the nucleus of NGC 1275 in degrees. As a result of the second
step, the accuracy of the determination of the coordinates of the
gamma-ray shower source increases by a factor of $\sim$10 compared
to the first step and it becomes possible to find the gamma-ray
intensity distribution in the source and its surroundings.

\begin{figure*}
\begin{center}
\includegraphics[width=2.8in]{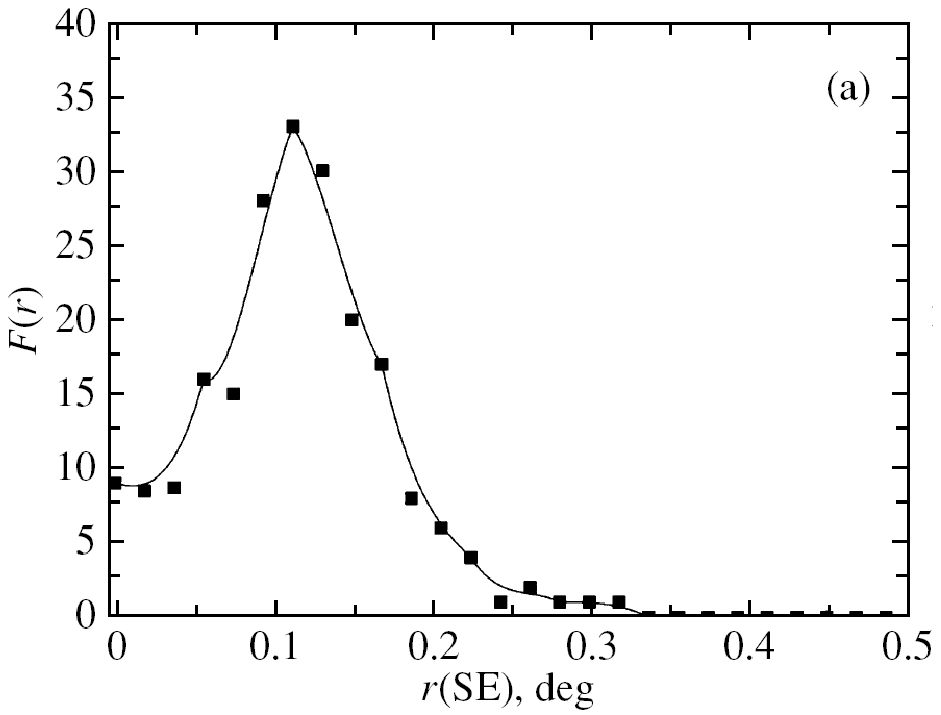}
\includegraphics[width=2.75in]{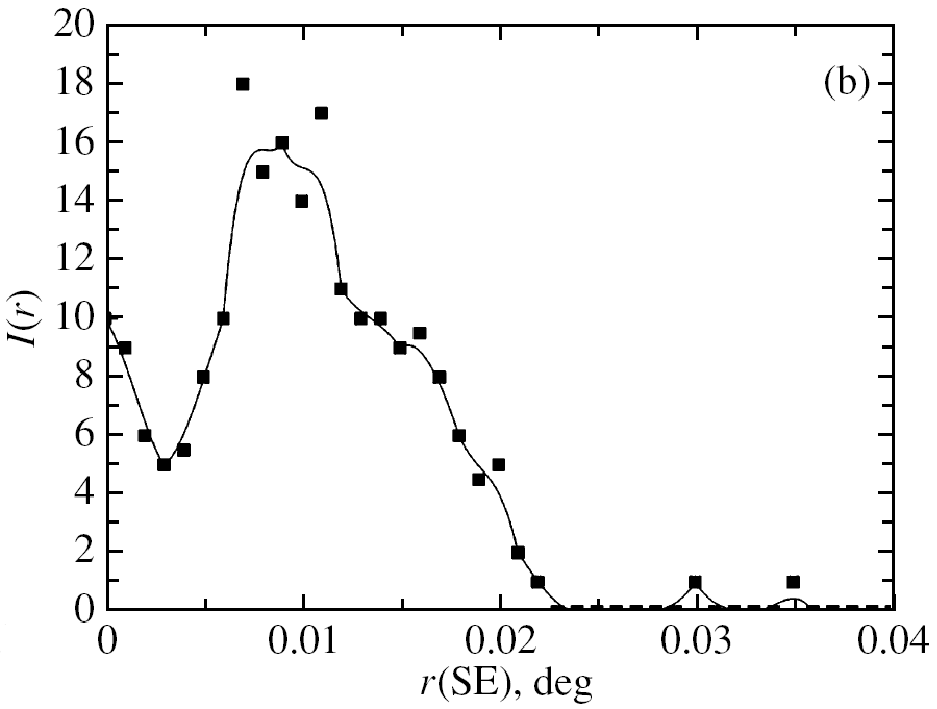}
\end{center}
\caption{(a) Gamma-ray intensity distribution obtained at the first
step; (b) gamma-ray intensity distribution in the source after the
correction at the second step.}
\end{figure*}
\begin{figure*}
\begin{center}
\includegraphics[width=5.5in]{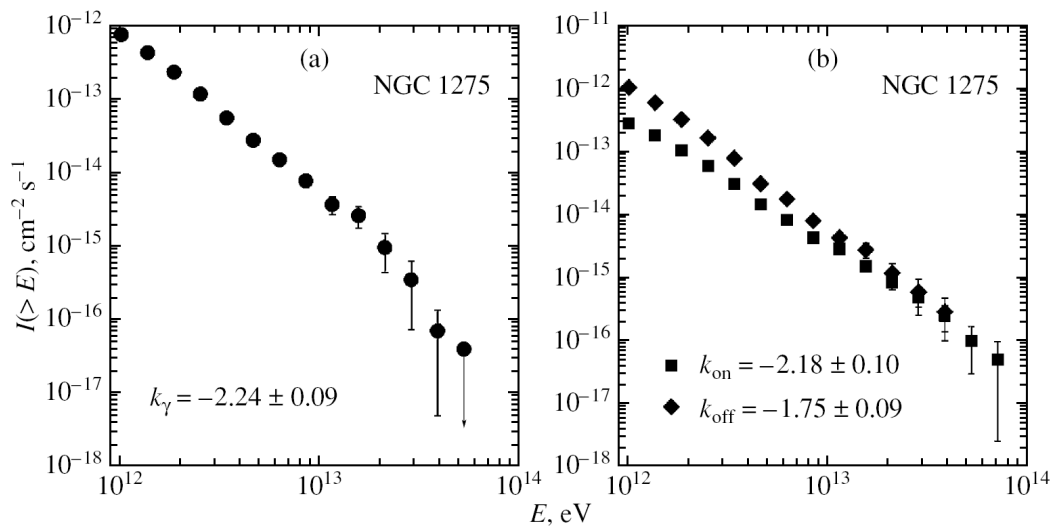}
\end{center}
\caption{(a) Gamma-ray spectrum of NGC1275 with a power-law index
$k_\gamma = -2.24\pm0.09$; (b) the spectrum of the events from NGC
1275 that passed the selection criteria without subtracting the
background with $k_{on} = -2.18 \pm 0.10$ and the spectrum of the
background events observed simultaneously with the source with
$k_{off} = -1.75 \pm 0.09$.}
\end{figure*}

\section*{NGC 1275 AT VERY HIGH ENERGIES}

Metagalactic sources of very high energy gamma-rays have been
searched for in the SHALON experiment from the very beginning of its
operation (Sinitsyna 1995, 1996). In 1996, the observations with the
SHALON mirror Cherenkov telescope revealed a new metagalactic source
of gamma-ray emission at very high energies $E >$ 800 GeV (Sinitsyna
et al. 1998; Sinitsyna 2000) (Figs. 8 and 9). The position of
emission source detected in our experiment coincides in its
coordinates with the Seyfert galaxy NGC 1275 (Sinitsyna 1997, 2000,
2006; Sinitsyna et al. 1998, 2003, 2007, 2009; Nikolsky and
Sinitsyna 2004). NGC 1275 was observed by the SHANON telescope for
271.2 h in different years (from 1996 to 2012) during the clear
moonless nights at zenith angles from 3$^\circ$ to 33$^\circ$. The
observations were performed using the standard (for SHALON)
technique of obtaining information about the cosmic-ray background
and gamma-ray-initiated showers in the same observing session
(Sinitsyna 1996; V.G. Sinitsyna and V.Yu. Sinitsyna 2011a, 2011b).
Gammaray emission from NGC 1275 was detected by the SHALON telescope
at energies above 800 GeV at the 31.4$\sigma$ confidence level
determined according to Li and Ma (1983). The average integral flux
at energies above 800 GeV for NGC 1275 is $I_{NGC 1275} = (7.8 \pm
0.5)\times 10^{-13}$ cm$^{-2}$ s$^{-1}$ (Fig. 8).

\begin{figure*}
\begin{center}
\includegraphics[width=6.1in]{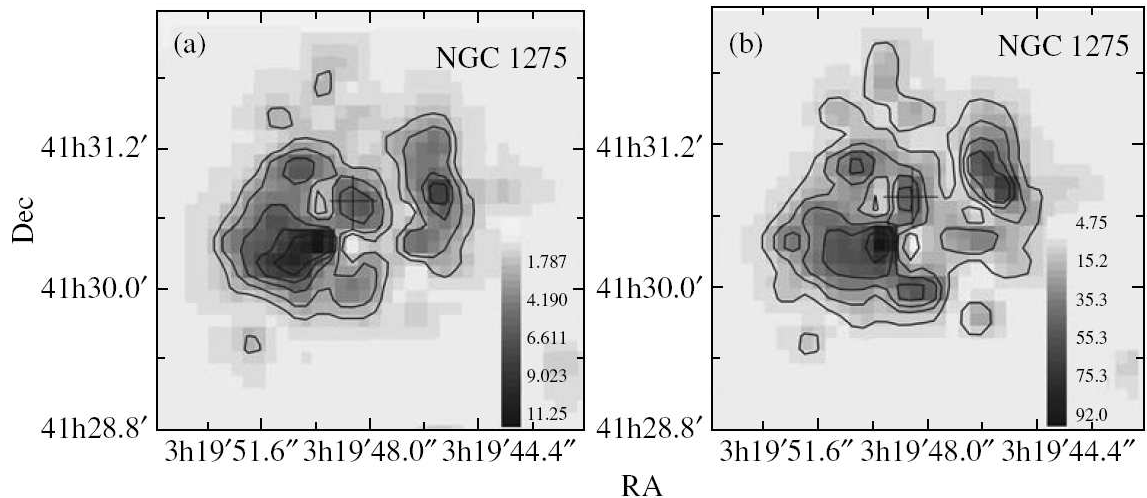}
\end{center}
\caption{(a) SHALON image of the gamma-ray source NGC 1275 at
energies $>$0.8 TeV; (b) the energy image of NGC 1275.}
\end{figure*}

Figure 8b presents the spectra of the $on$ and $off$ events needed
to extract the gamma-ray spectrum from NGC 1275. The gamma-ray
spectrum of NGC 1275 is obtained by subtracting the spectrum of the
background events recorded simultaneously with the source's
observations, $I_{off} \propto E^{k_{off}}$, from the spectrum of
the events arrived from the source position, $I_{on} \propto
E^{k_{on}}$. The integral spectrum of the $off$ events was
constructed from statistics larger than that of $on$ events by a
factor of 4, because it is an averaging over the background taken
from four regions far away from the source but equal in angular size
to the source's region. The gamma-ray energy spectrum in the
observed energy range above 0.8 TeV is well described by a power law
$F(E_0 > 0.8$ TeV$) \propto E^{k_{\gamma}}$, where $k_{\gamma} =
-2.24 \pm 0.09$ (see Fig. 8a; see the Appendix).

Figure 9 presents the source's image at TeV energies and its energy
image by SHALON experiment. The color scale in Fig. 9a on the left
is in units of the excess above the minimum detected signal for the
energy image (in TeV).

\begin{figure*}
\begin{center}
\includegraphics[width=3.5in]{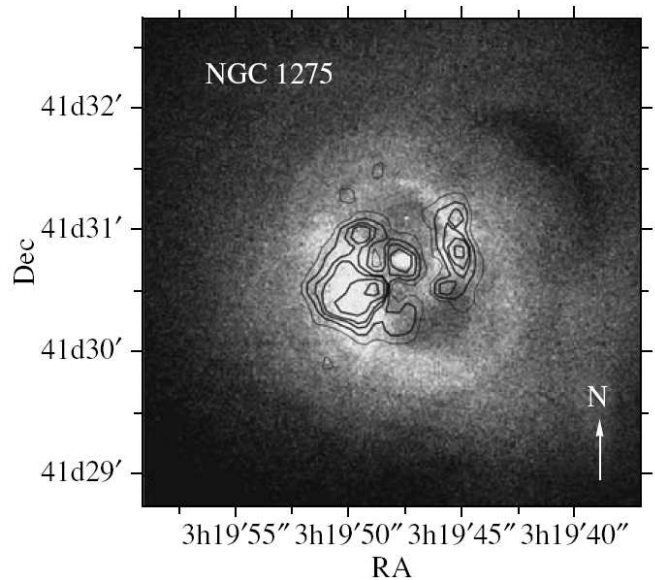}
\end{center}
\caption{Chandra X-ray (1.5 –- 3.5 keV) image of NGC 1275 (Fabian et
al. 2000); the contours indicate the SHALON image of NGC 1275 in the
energy range 800 GeV –- 40 TeV.}
\end{figure*}
\begin{figure}[b!]
\begin{center}
\includegraphics[width=3.1in]{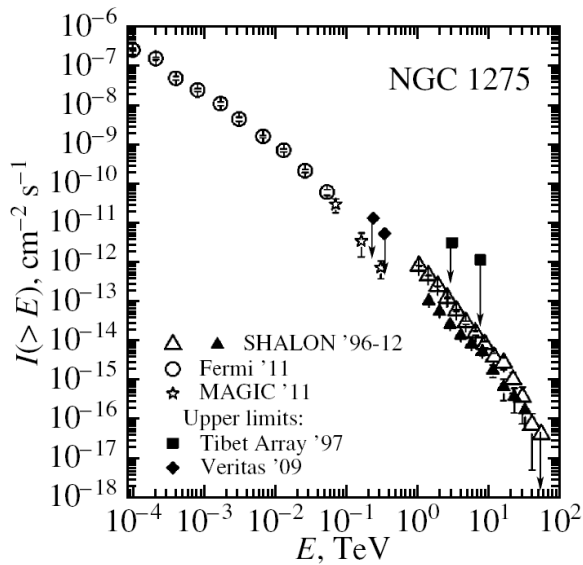}
\end{center}
\caption{Integral spectrum of high and very high energy gamma rays
from NGC 1275 obtained by SHALON in comparison with the data from
the Fermi LAT satellite telescope, the MAGIC and VERITAS mirror
Cherenkov telescopes, and the scintillation Tibet Array.}
\end{figure}

Possible correlations between the emission regions of TeV gamma rays
and low-energy (radio and X-ray) photons should be established to
elucidate the mechanisms of the generation of very high energy
emission in the source and to test the models describing them.
Figure 1 shows a ROSAT X-ray image of NGC 1275 (black-and-white
scale); the contours represent the source's radio structure from VLA
radio observations. The radio and X-ray emission maxima coincide
with the AGN NGC 1275. At the same time, the X-ray emission
disappears almost completely near the bright areas of the radio
components located in the north and the south symmetrically relative
to the core (B\"ohringer et al. 1993). We also combined the SHALON-1
(0.8 -- 40 TeV) and Chandra (1.5 -- 3.5 keV X-ray) images. Figure 10
(black-and-white scale) presents a Chandra X-ray (1.5 -- 3.5 keV)
image for the central part of the Perseus cluster centered on NGC
1275 with a size of $\sim$5.5 arcmin (Fabian et al. 2000). In the
X-ray energy range, the core of the Perseus cluster, on the whole,
appears as a clear circularly symmetric structure with a distinct
maximum on NGC 1275.

The clearly seen dimming in X-ray flux, along with the dip NW of the
center, known from the 1979 Einstein observations (Fabian et al.
1981), correlates with the components of the extended double radio
structure 3C 84 (Fig. 1). These dips are surrounded by bright (at
energies 1.5 –- 3.5 keV) arc regions from the north and the south.
The simplest interpretation is that the intense emission from these
rims comes from the shells surrounding the radio lobes (Fabian et
al. 2000). A bright emission spot is also observed to the east.

The emission regions of very high energy gamma rays observed by
SHALON from NGC 1275 have a structure similar to that described
above (see Fabian et al. 2000) and well correlates with the photon
emission regions in the energy range 1.5 –- 3.5 keV (Fig. 10). A
correlation of the emission with energies 0.8 –- 40 TeV (Sinitsyna
et al. 2003, 2007, 2009) and the X-ray emission in the range 0.3 –-
7 keV (Fabian et al. 2000) was also found. Thus, the TeV gamma-ray
emission recorded by SHALON from NGC 1275 has an extended structure
with a distinct core centered at the source's position.

To analyze the emission related to this core, we additionally
identified the emission component corresponding to the central
region of NGC 1275 with a size of 32". The emission from the central
region of NGC 1275 was detected at energies above 0.8 TeV at a
13.5$\sigma$ confidence level determined by the Li\&Ma method (Li
and Ma 1983) with a average integral flux $I(>800$ GeV$) = (3.26 \pm
0.30) \times 10^{-13}$ cm$^{-2}$ s$^{-1}$. The gamma-ray energy
spectrum of the central component in the entire energy range from
0.8 to 40 TeV is well described by a power law with an exponential
cutoff, $I(>E_{\gamma}) = (2.92\pm0.11)\times10^{-13}\times
E_{\gamma}^{-1.55\pm 0.10}\times$ exp$(-E_{\gamma}/$10 TeV$)$
cm$^{-2}$ s$^{-1}$. (see Fig. 11, the black triangles). When the
source's spectrum in the energy range 0.8 -- 20 TeV is described by
a simple power law $F(E_0 > 0.8$ TeV$) \propto E^{k_{\gamma}}$, the
spectral index is $k_{\gamma} = -1.92 \pm 0.11$. The SHALON spectrum
corresponding to the emission from the central region of NGC 1275 is
represented in Fig. 11 by the black triangles.

Recently, the AGN NGC 1275 was also recorded by the MAGIC
ground-based mirror Cherenkov telescope at energies above 100 GeV in
the 2010 –- 2011 observations (Al\'eksic et al. 2012). Figure 11
compares the integral gamma-ray spectrum of NGC1275 and its central
region obtained from SHALON data (1996 –- 2012) with the Fermi LAT
(2009 –- 2011) (Brown and Adams 2011) and MAGIC (2010 –- 2011)
(Al\'eksic et al. 2012) experimental data.

\section*{VARIABILITY OF THE GAMMA-RAY EMISSION FROM NGC 1275}

Reliably revealing flares and their duration in long-term
observations with mirror Cherenkov telescopes is complicated by the
fact that the technique makes a continuous tracking of the source
impossible, because it requires such conditions as moonless nights,
which already creates a gap in the data for more than ten days; an
ideal atmosphere without clouds and haze and, in addition, the
source's passage at a distance of no more than 35$^\circ$. from
zenith are needed, because the influence of a change in atmospheric
thickness should be minimal.

Nevertheless, revealing correlations between the emissions in
different energy ranges, comparing the emission regions, and, in
particular, the detection of the flux changes remains necessary,
because it makes it possible to judge the nature of the source, its
evolution, and the emission generation mechanisms in various
objects.

The observed gamma-ray flux variations, on average, do not exceed
20\% of $(7.8 \pm 0.5) \times 10^{-13}$ cm$^{-2}$ s$^{-1}$. The
SHALON mirror Cherenkov telescope has detected three short-time
(within five days) increases and one decrease of the very high
energy gamma-ray flux in the entire time of observations of NGC
1275. Given these variations, the flux decrease below the average
was recorded in 1999 and the integral flux was $(4.7 \pm 1.3)\times
10^{-13}$ cm$^{-2}$ s$^{-1}$. The increases were detected in late
January 2001, late November–early December 2005, and late October
2009. The fluxes in these periods were $(21.2\pm 7.5)\times
10^{-13}$, $(35.5\pm 12.4)\times10^{-13}$, and $(23.4 \pm 4.5)\times
10^{-13}$ cm$^{-2}$ s$^{-1}$, respectively. The duration of the flux
increase in October 2009 was 3 days. No intervals of flux increase
were found in 2001 and 2005, because the observations were
interrupted due to weather conditions in both cases.

To reveal possible correlations of the emissions in various energy
ranges, including those at high and very high energies, we compared
the NGC 1275 gamma-ray fluxes by SHALON in the periods when the
observations were simultaneous with the ones by the Fermi LAT
experiment. The published Fermi LAT data were obtained from August
4, 2008, to September 30, 2010 (Brown and Adams 2011). The SHALON
observations of NGC 1275 were performed in November 2008 with a
break for the Moon's time, October 2009, and mid-November-early
December 2010. In this time, only one gamma-ray flux increase to
$(23.4 \pm 4.5)\times 10^{-13}$ cm$^{-2}$ s$^{-1}$ was detected in
the period of October 18 –- 20, 2009. These periods of SHALON
observations do not coincide with the times of the main flares
observed at Fermi LAT (Brown and Adams 2011). A slight local flux
increase can be seen in the period of mid-October 2009 (Brown and
Adams 2011), which corresponds to the above-mentioned gamma-ray flux
increase observed by SHALON.

\section*{NGC 1275 AS A POINT AND EXTENDED SOURCE}

As has been pointed out above, the Perseus cluster of galaxies with
the central galaxy NGC 1275 is ideally suitable both for studying
the physics of relativistic jets from AGNs and for revealing the
feedback role of the central galaxy. Evidence for the latter was
obtained in ROSAT and Chandra observations at low energies, from
which shells of hot gas and cavities that spatially coincide with
the radio structures originating in the central, active part of the
AGN can be seen (Fig. 1 and B\"ohringer et al. 1993; Fabian et al.
2000, 2006; Churazov et al. 2000). The observational data for NGC
1275 at energies 800 GeV -– 40 TeV, namely the images of the galaxy
and its surroundings (Figs. 9 and 10), as well as the flux
variations suggest that the TeV gamma-ray emission in these regions
is produced by a number of processes.
\begin{figure}[t!]
\begin{center}
\includegraphics[width=3.4in]{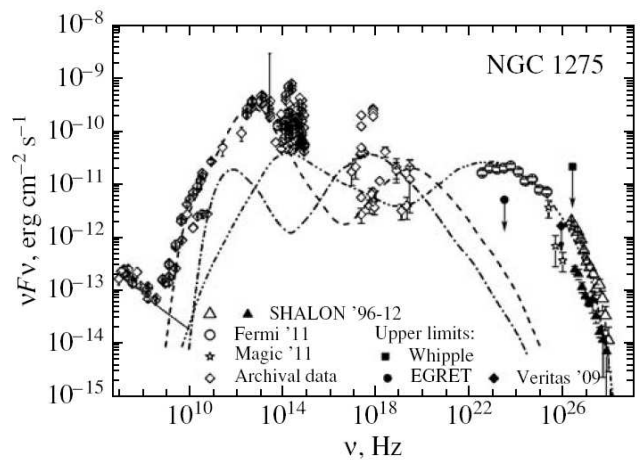}
\end{center}
\caption{Spectral energy distribution of the gamma-ray emission from
NGC 1275. $\triangle$, $\blacktriangle$ -- represent the data from
the SHALON ground-based Cherenkov telescope. $\bigcirc$ -- represent
the Fermi LAT and Magic data. The arrows indicate the upper limits
from EGRET, Whipple, and VERITAS data (see the text). The dashed,
dash–dotted, and dash–dotted with two dots curves indicate the
spectral energy distributions of NGC1275 obtained in the CM model
(Colafrancesco et al. 2010).}
\end{figure}

The extended structure around NGC 1275 (Fig. 9) that spatially
coincides with the X-ray emission regions (Fig. 10) can be produced
by mechanisms related to the generation of an X-ray structure
(Fabian et al. 2000, 2006; Churazov et al. 2000). The brightness
distribution of the X-ray emission and the observed TeV emission
shows a sharp increase in intensity immediately outside the bubbles
blown by the central black hole and visible in the radio band. This
suggests that the X-ray-generating particles are swept up from the
region of the radio lobes under the pressure of cosmic rays and
magnetic fields generated in the jets at the center of NGC 1275
(Fabian et al. 2006; Churazov et al. 2000). The structures visible
in TeV gamma rays are formed through the interaction of very high
energy cosmic rays with the gas inside the Perseus cluster and
interstellar gas heating at the boundary of the bubbles blown by the
central black hole in NGC 1275.

The presence of emission in the energy range 1¨C40 TeV from a
central region of $\sim$32" in size around the nucleus of NGC 1275
(see Fig. 11, the black triangles) and the short-time flux
variability point to the origin of the very high energy emission as
a result of the generation of jets ejected by the central
supermassive black hole of NGC 1275. The multifrequency spectral
energy distribution for the nucleus of NGC 1275, up to high and very
high energies, was described in the CM model (Colafrancesco et al.
2010) and is a composition of the components of inverse Compton
scattering of the intrinsic synchrotron radiation from relativistic
electrons (synchrotron self-Compton) of three separate plasma blobs
ejected from the inner regions of the NGC 1275 nucleus (Fig. 12, the
dashed, dash– dotted, and dash–dotted with two dots curves). The
available Fermi LAT data at high energies and the SHALON
observations at very high energies in a region $<$32" around NGC1275
are described in terms of this model with one of the components
produce synchrotron self-Compton emission of the relativistic jets
from the nucleus itself (Fig. 12, the dash–dotted with two dots
curve).

\section*{CONCLUSION}
The cluster of galaxies in Perseus, along with other clusters, have
long been considered as possible candidates for the sources of high
and very high energy gamma-ray emission generated by various
mechanisms. Long-term studies of the central galaxy in the cluster,
NGC 1275, are being carried out in the SHALON experiment. We
presented the results of fifteen-year-long observations of the AGN
NGC 1275 at energies 800 GeV –- 40 TeV discovered by the SHALON
telescope in 1996 (Sinitsyna 1997, 2000; Sinitsyna et al. 1998,
2003, 2006, 2007, 2009; Nikolsky and Sinitsyna 2004). The data
obtained at very high energies, namely the images of the galaxy and
its surroundings, and the flux variability indicate that the TeV
gamma-ray emission is generated by a number of processes: in
particular, part of this emission is generated by relativistic jets
in the nucleus of NGC 1275 itself. Whereas, the presence of an
extended structure around NGC 1275 is evidence of the interaction of
cosmic rays and magnetic fields generated in the jets at the
galactic center with the gas of the Perseus cluster.

\vspace{8mm}

\section*{\it APPENDIX}

\begin{table}[b!]
\renewcommand{\arraystretch}{1.4}
\vspace{6mm} \centering { Flux measurements in each energy interval 
}\label{tabl1} \vspace{3mm}
\begin{tabular}{c|c|c|c} \hline\hline
E, TeV   & Excess      & $\sigma$,  &   $dF/dE$,    \\
         &             & (Li\&Ma)   &   cm$^{-2}$s$^{-1}$TeV$^{-1}$    \\
\hline
0,904        & 39      &  8,3   & $(9,59\pm 1,53)\times10^{-13}$     \\
1,22         & 54      & 10,6   & $(4,45\pm 0,60)\times10^{-13}$     \\
1,66         & 41      &  8,9   & $(1,39\pm 0,22)\times10^{-13}$     \\
2,25         & 47      &  9,8   & $(7,52\pm 1,01)\times10^{-14}$     \\
3,05         & 64      & 14,5   & $(5,26\pm 0,65)\times10^{-14}$       \\
4,13         & 34      & 10,3   & $(1,51\pm 0,26)\times10^{-14}$     \\
5,60         & 19      &  8,2   & $(4,76\pm 0,91)\times10^{-15}$\\
7,59         & 18      &  8,1   & $(2,63\pm 0,62)\times10^{-15}$\\
10,2         & 6       &  4,3   & $(4,44\pm 1,98)\times10^{-16}$\\
13,9         & 10      &  5,9   & $(4,83\pm 1,61)\times10^{-16}$\\
18,9         & 5       &  4,0   & $(1,32\pm 6,62)\times10^{-16}$\\
25,6         & 4       &  3,8   & $(6,27\pm 3,62)\times10^{-17}$\\
34,7         & 2       &  2,7   & $(2,47\pm 1,75)\times10^{-17}$\\
47,0         & 1       &  1,5   & $(1,75\pm 1,00)\times10^{-17}$ \\
\hline \multicolumn{4}{l}{}\\ 
\end{tabular}
\end{table}

\begin{figure}[b!]
\begin{center}
\includegraphics[width=2.9in]{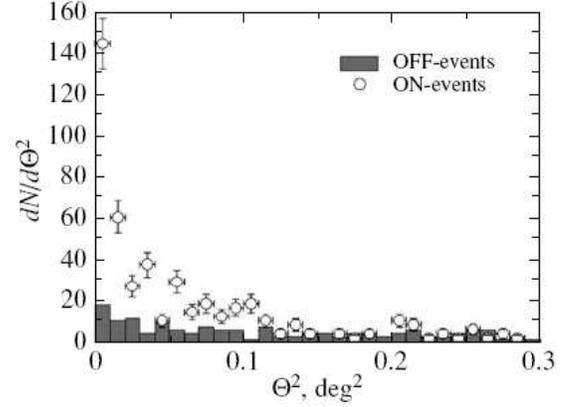}
\end{center}
\caption{Distribution of $\Theta^2$ for the $on$ and $off$ events
recorded in the SHALON observations of NGC 1275.}
\end{figure}
\begin{figure}[b!]
\begin{center}
\includegraphics[width=2.8in]{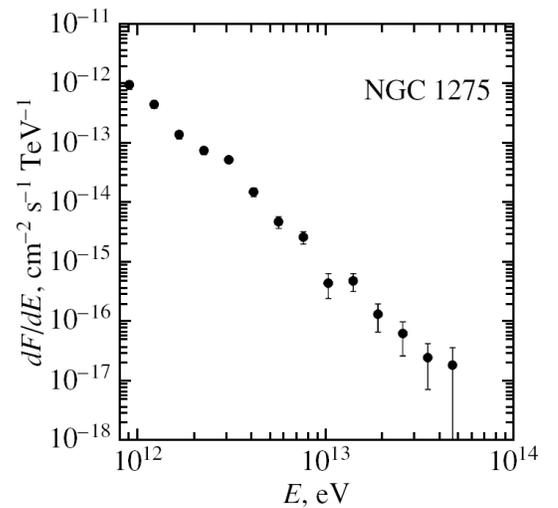}
\end{center}
\caption{Differential spectrum of NGC 1275 from SHALON data.}
\end{figure}

Figure 13 presents the distribution in $\Theta^2$ for the signal
($on$) and background ($off$) events recorded during the
observations of NGC 1275 in the SHALON experiment in 271.2 h.
$\Theta^2$ is the distance between the source's position and the
shower direction to the source reconstructed in the experiment. The
observed excess corresponds to 31.4$\sigma$ determined according to
Li and Ma (1983).

The table gives the signal excess above the background for each
energy interval of the differential spectrum, the signal detection
confidence level according to Li and Ma (1983) in each interval, and
the differential flux. The differential spectrum of NGC 1275 is
presented in Fig.~14.

\end{document}